# Resistless EUV lithography: photon-induced oxide patterning on silicon

Li-Ting Tseng[1†], Prajith Karadan[1], Dimitrios Kazazis[1]*, Procopios C. Constantinou[1], Taylor J. Z. Stock[2,3], Neil J. Curson[2,3], Steven R. Schofield[2,4], Matthias Muntwiler[1], Gabriel Aeppli[1,5,6] and Yasin Ekinci[1]

[1] Paul Scherrer Institute, 5232 Villigen PSI, Switzerland

[2] London Centre for Nanotechnology, University College London, London, WC1H 0AH, UK

[3] Department of Electronic and Electrical Engineering, University College London, London, WC1E 7JE, UK

[4] Department of Physics and Astronomy, University College London, London, WC1E 6BT, UK

[5] Laboratory for Solid State Physics and Quantum Center, ETH Zürich, 8093 Zürich, Switzerland

[6] Institut de Physique, EPFL, 1015 Lausanne, Switzerland

*Dimitrios Kazazis, Email: dimitrios.kazazis@psi.ch

## Abstract

In this work, we show the feasibility of extreme ultraviolet (EUV) patterning on an HF-treated Si(100) surface in the absence of a photoresist. EUV lithography is the leading lithography technique in semiconductor manufacturing due to its high resolution and throughput, but future progress in resolution can be hampered because of the inherent limitations of the resists. We show that EUV photons can induce surface reactions on a partially H-terminated Si surface and assist the growth of an oxide layer, which serves as an etch mask. This mechanism is different from the H-desorption in scanning tunneling microscopy-based lithography. We achieve $SiO_2$/Si gratings with 75 nm half-pitch and 31 nm height, demonstrating the efficacy of the method and the feasibility of patterning with EUV lithography without the use of a photoresist. Further development of the resistless EUV lithography method can be a viable approach to nm-scale lithography by overcoming the inherent resolution and roughness limitations of photoresist materials.

## Teaser

Performing EUV lithography without the use of a photoresist to enable nanometer-scale patterning in future technology nodes.

## Introduction

Pushing the resolution limits of photolithography has been of paramount importance in the past decades to keep up with transistor downscaling, following Moore's law (*1*). In photolithography, the resolution is ultimately limited by wavelength. Whereas deep ultraviolet (DUV) lithography with 193 nm wavelength has been the workhorse of semiconductor manufacturing in the last decade, extreme ultraviolet (EUV) lithography has recently been introduced in semiconductor manufacturing at the 7 nm node, owing to its short wavelength of 13.5 nm and high throughput (*2*), enabling further downscaling of semiconductor devices for future technology nodes. Nevertheless, as feature sizes continue to decrease, the limitations posed by the available photoresists become critical and may eventually limit further progress (*3*).

The resolution limitation and intrinsic line-edge roughness of the photoresists come mainly from stochastic material properties and processes (*4*). Typically, resists are composed of large molecules and multiple components which have compositional and density fluctuations at the atomic scale. Moreover, diffusion, in particular for chemically-amplified resists (CARs) (*5*), and the stochastic nature of the development after the exposure (*6*) are important factors that induce line-edge roughness and eventually limit the ultimate resolution of the photoresist. Recently, there have been promising attempts to scale down the resist thickness by patterning self-assembled monolayers with EUV light, combined with selective growth (*7*) or selective-area deposition (*8*).

On the other hand, H-terminated Si has widely been used in the fabrication of quantum devices (*9-12*). At the atomic scale, the patterning of H-terminated Si can be achieved by desorbing H atoms on the Si surface using scanning tunneling microscopy (STM) (*13, 14*), followed by selective-area doping (*15*) or selective atomic layer deposition (ALD) (*16-18*) or by inducing localized oxidation (*19*). These methods have been a clear demonstration of top-down lithography with nm-scale resolution. It has also been demonstrated that selective H desorption (from hydrogen-terminated Si) can be achieved by electron-beam lithography (EBL) (*20*) and vacuum ultraviolet (VUV) photons (*21*). However, all the aforementioned lithographic techniques are direct-write methods, suffering from low throughput (*22*). Kramer *et al.* have demonstrated ultraviolet (UV) photolithography patterning directly on an H-passivated Si substrate and obtained a grating with a half-pitch of 500 nm (*23*).

Inspired by these earlier efforts, in this work, we study the potential of direct patterning on partially H-terminated Si, instead of bulk photoresists, for use in the future nodes of EUV lithography by overcoming the limitations of photoresist materials. This work is not only relevant to advanced semiconductor manufacturing but also for the development of quantum devices at a large scale, limiting the use of time-consuming STM lithography only to the patterning of ultrafine, non-periodic structures. We show high-resolution resistless EUV lithography patterning through EUV exposure on HF-treated Si surfaces using the achromatic Talbot lithography (ATL) technique, which will be described later (*24*). The partially H-terminated Si (100) wafers, prepared with HF treatment, are exposed to EUV light under high vacuum resulting in the formation of a dense oxide layer at the exposed area. The oxide nanopattern is, in turn, transferred to the Si substrate by etching in a tetramethylammonium hydroxide (TMAH) solution, resulting in $SiO_2$/Si nanowires of half-pitch 75 nm. Extensive studies on the exposed samples using X-ray photoelectron spectroscopy (XPS) reveal a mechanism that is different from STM-based H-depassivation lithography. XPS results demonstrate that the exposure to EUV light promotes the formation of Si oxides leading to stable $SiO_2$ at high doses. At a dose below 68 $J/cm^2$, the formation of suboxides dominates, whereas, the growth of $SiO_2$ prevails at higher doses. We believe that by extending this patterning approach of high-resolution and large-area resistless patterning to other atomically thin layers, for example, Cl or I terminated Si or Si with OH/$SiO_2$ terminations, one could potentially overcome the resolution limitations of conventional photoresists for future lithography nodes and by combining it, for example, with selective-area doping one could also enable the upscaling of devices for quantum computing.

## Results

### Resistless lithography and process optimization

A schematic diagram showing our concept of EUV resistless lithography is shown in Fig. 1. The scheme is as follows: A p-type Si (100) wafer is immersed into a buffered HF solution to remove its native oxide layer, leaving behind an H-passivated Si surface (*25*). The wafer is immediately loaded into the EUV interference lithography vacuum chamber located at the XIL-II beamline of the Swiss light source (SLS) (*26*). We note that there is a half-hour delay between the surface preparation and pump-down of the lithography tool and during this time the samples are packed in a nitrogen-flushed transfer box and exposed to ambient air for about 5 minutes. Therefore, a partial re-oxidation of the Si surface is expected. However we expect that the thickness of any oxide on the surface is on the angstrom scale as the sample is exposed to air for less than 1 h before being introduced in the EUV chamber (*27*). The samples are selectively exposed to EUV light, either through a square aperture to obtain a contrast curve or using special EUV achromatic Talbot lithography masks to obtain high resolution and dense line/space patterns. The EUV exposure on the partially oxidized surface promotes the formation of silicon oxide in the exposed region, due to the residual water on the sample surface, arising since the wafer is rinsed in water after the HF treatment, or in the chamber background. The presence of water and oxygen in the chamber is, indeed, confirmed by residual gas analysis. The generation of secondary electrons due to the EUV absorption plays a key role in the dissociation of absorbed water, which promotes the formation of silicon oxide. The Si surface is subsequently treated in TMAH solution, which has a high etch selectivity of Si over its oxides (*28*). In this way, the oxide pattern can be transferred to the Si substrate.

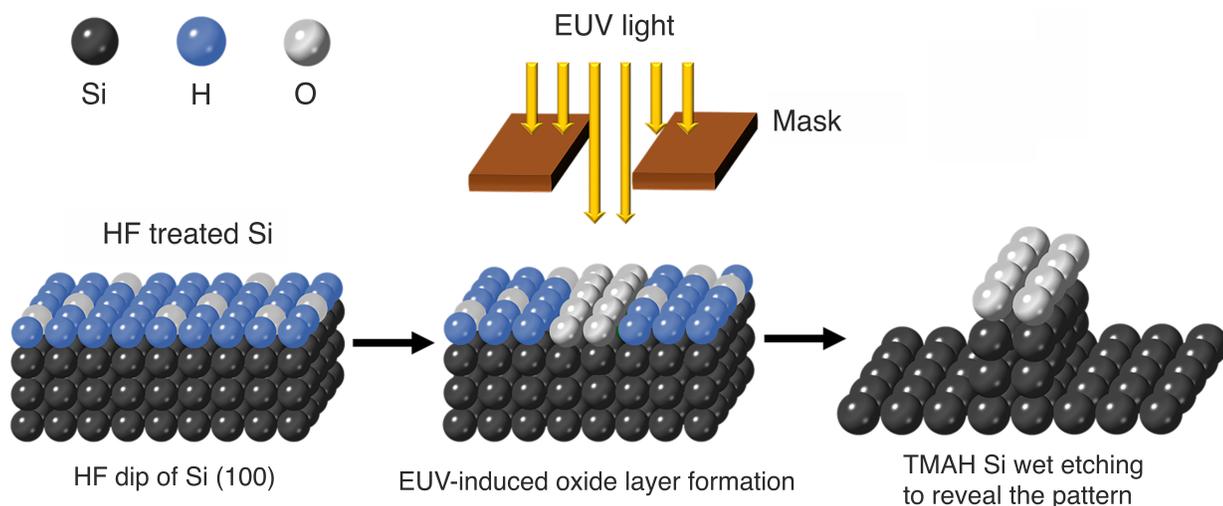

**Fig. 1. Schematic of the EUV resistless patterning process of an HF-treated Si (100) surface.** A Si wafer is treated with an HF solution to create a mostly H-passivated surface. The wafer is subsequently inserted in a high-vacuum chamber and exposed to EUV light. This promotes the oxide formation on the surface. The wafer is finally treated with TMAH to reveal the pattern by etching into Si using the oxide as a mask.

We start the study by exploring the experimental parameters of exposure and etching. Figure 2A shows the simulated attenuation length of EUV photons in Si as a function of their energy (*29*). The exposure mechanism at EUV wavelengths is governed by secondary electrons, which can travel a few nanometers to induce chemical changes (*30-32*). Since the major part of the absorption of the photons occurs in the bulk, we can expect that the electrons generated by the photons absorbed in bulk Si close to the surface should be able to diffuse to the surface and induce chemical changes there. From Fig. 2A, we see that there is a strong absorption edge at around 100 eV, corresponding to the Si 2p core-level excitation. Therefore, we conducted our optimization experiments at an energy of 102 eV, slightly above the Si 2p absorption edge, where we expect an increased photoelectron yield due to the higher photon absorption. To optimize our process, we exposed HF-treated Si wafers at an energy of 102 eV with a selected range of doses through a 0.5×0.5 mm$^2$ aperture (open-frame exposures) and performed etching with different TMAH solutions, temperatures, and etch times. The results of our optimization process are summarized in Fig. 2B, which shows the measured etch depth of Si as a function of EUV dose at a photon energy of 102 eV for three different TMAH solutions. The etching time was 5 s and the temperature of the solution was kept at 85 °C. These curves are similar to the contrast curves of photoresists (33), which show the remaining photoresist thickness after development as a function of the exposure dose. In our case, the dose-dependent thickness corresponds to the depth of the Si etch in TMAH, instead of the remaining resist. These dose-response curves show a negative-tone behavior (higher etch depth for higher doses), as expected from the patterning mechanism. The 25 wt% TMAH solution gives a higher contrast but shows very low sensitivity (high dose to obtain a pattern). A more dilute (10 wt%) TMAH solution increases the sensitivity by more than an order of magnitude, but also degrades the contrast of the curve as shown in Fig. 2B. By adding 30 vol% of isopropyl alcohol (IPA) to the 10 wt% TMAH solution, the patterns appear at even lower doses. Adding IPA in TMAH for Si etching also enhances the anisotropic etching behavior along the [100] direction and smoothens the etched Si surface (28, 34). This makes it more suitable for high-resolution patterning.

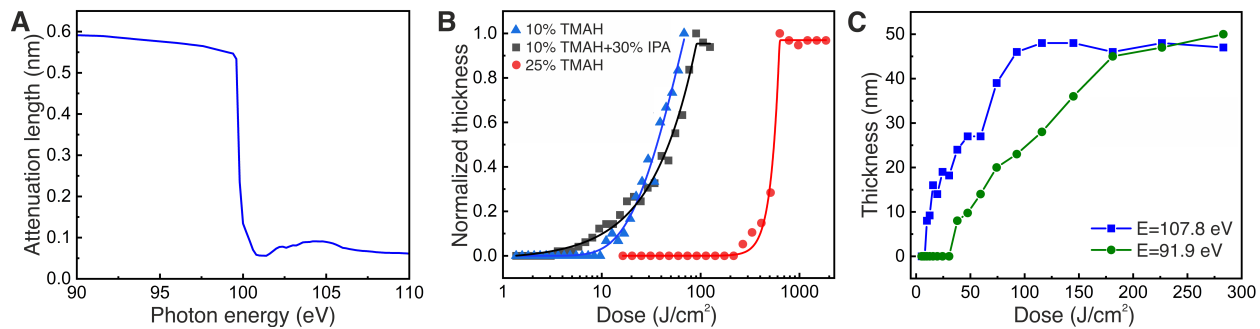

**Fig. 2. Absorption edge of Si and EUV exposures above and below it.** (**A**) Simulated X-ray attenuation length as a function of photon energy in Si. (**B**) A plot of normalized Si etch thickness versus EUV exposure dose with different TMAH concentrations at an energy of 102 eV, above the Si 2p absorption edge. The etching was performed at 85 °C for 5 s. (**C**) Measured Si etch thickness as a function of EUV exposure dose at two different photon energies below and above the Si absorption edge. The etching was, again, performed at 85 °C for 5 s.

However, for this patterning process to be attractive and relevant for industrial integration, it needs to be applied at the semiconductor manufacturing EUV wavelength of λ=13.5 nm, corresponding to a photon energy of E=91.9 eV. We note that this photon energy is below the 2p absorption edge of Si, shown in Fig. 2A, and that bulk Si is relatively transparent at this wavelength. To investigate the mechanism of the chemical reactions, we carried out exposures at two photon energies, below and above the Si absorption edge at 91.9 and 107.8 eV, respectively. We performed the experiments on the same Si wafer, under identical conditions, and with the same post-exposure treatment. As shown in Fig. 2C, the Si etch depth is measured and plotted as a function of the EUV exposure dose. Similarly to Fig. 2B, the curves show a negative-tone behavior and the Si etch depth saturates after a certain dose, which we refer to as the "dose-to-pattern" in analogy to the "dose-to-gel" for a negative tone resist. For both energies, the Si etch depth reaches the same maximum at the dose-to-pattern value, which is about a factor 2 lower for E=107.8 eV compared to E=91.9 eV. We attribute this to the increased absorption cross section of Si and relatively higher photoelectron yield at 107.8 eV. At 91.9 eV, the Si 2p core-level is not excited, so fewer photoelectrons are generated in comparison to 107.8 eV. At 107.8 eV, Si 2p core-level excitation and thereby the corresponding LVV Auger relaxation leads to the generation of more secondary electrons near the surface. The high rate of oxide formation at 107.8 eV is due to the availability of more secondary electrons at the surface due to the increased absorption and photoelectron yield. However, it must be noted that the Si etch thickness saturates almost at the same point for both energies. This suggests that the photon energy has an effect on the rate of oxide formation. Moreover, it is a striking observation that it is possible to etch 50 nm into Si simply by a surface modification. This is a first indication that the actual patterning mechanism is more complex than originally thought. The secondary electrons at the surface play a role in dissociating adsorbed water and enabling the oxide layer formation. However, in addition, the large depths of the patterning we have achieved suggest an important role of the secondary electrons in the formation of more stable, homogeneous, and thicker oxides.

For this work to be relevant for industrial integration, the photon energy needs to be tuned to 91.9 eV, corresponding to a wavelength of 13.5 nm. Therefore, we have performed all further experiments, including the high-resolution patterning, at this energy and wavelength.

**Investigation of the patterning mechanism via XPS analysis**

To further investigate the exposure mechanism and the effect of EUV photons on the HF-treated Si (100) surface, and to understand how the EUV dose affects the oxidation process at the exposed Si surface, we used synchrotron radiation-based XPS. A synchrotron source has several advantages like high brilliance, high spectral resolution, and, most importantly, energy tunability. For comparison, we performed measurements on an untreated Si (100) wafer with native oxide on its surface, on an as-prepared HF-treated Si (100) wafer, and on an HF-treated Si surface exposed with EUV photons at an energy of E=91.9 eV. We conducted the experiments at three different EUV doses, a low dose of 6.8 J/cm$^2$, a medium dose of 68 J/cm$^2$, and a high dose of 204 J/cm$^2$. From Fig. 2C, we expect the low dose should correspond to no measured etched depth, the medium dose to a depth of 20 nm, and the high dose to a depth of 45 nm.

First, we carried out survey scans to investigate the composition of the Si surfaces. Figure 3 compares the survey scans of an untreated Si surface with its native oxide, an as-prepared HF-treated Si surface before any EUV exposure, and an HF-treated Si surface after EUV exposure, but at the unexposed part of the substrate. We use the latter as a control sample to monitor the effect of loading, pumping down, venting, and unloading from the EUV exposure chamber. As can be seen in Figs. 3A-C, Si, O, and C are detected in all samples. The carbon signal is due to slight contamination on the sample surface. We also note that small amounts of fluorine are also detected in the sample that had previously been loaded into the XIL-II exposure chamber (Fig. 3C). This is attributed to low level fluorine contamination in the XIL-II exposure chamber. An oxygen signal is also detected on the as-prepared HF-treated Si surface, indicating that it has already been slightly oxidized before the EUV exposure (Fig. 3B). However, the same O peak for the untreated Si surface is much higher (Fig. 3A). The comparison of the various peaks in the three samples can be visualized in Fig. 3D, which shows the peak-area ratios of O, F, and C over Si, calculated from the survey scans. These ratios indicate the relative composition of those three elements at the different Si surfaces. The results show that the oxygen signal is significantly lower in the HF-treated Si compared to the untreated Si surface, indicating a successful removal of the native oxide and passivation, although partial oxidation of the surface is unavoidable during the sample transfer. We also observe that the O/Si ratio is lower in the control sample with respect to the HF-treated one. We attribute this to the fact that the control sample underwent one additional loading, pumping, and unloading cycle in the EUV chamber compared to the HF-etched sample, which could help desorb residual water adsorbed on the surface during the rinsing step.

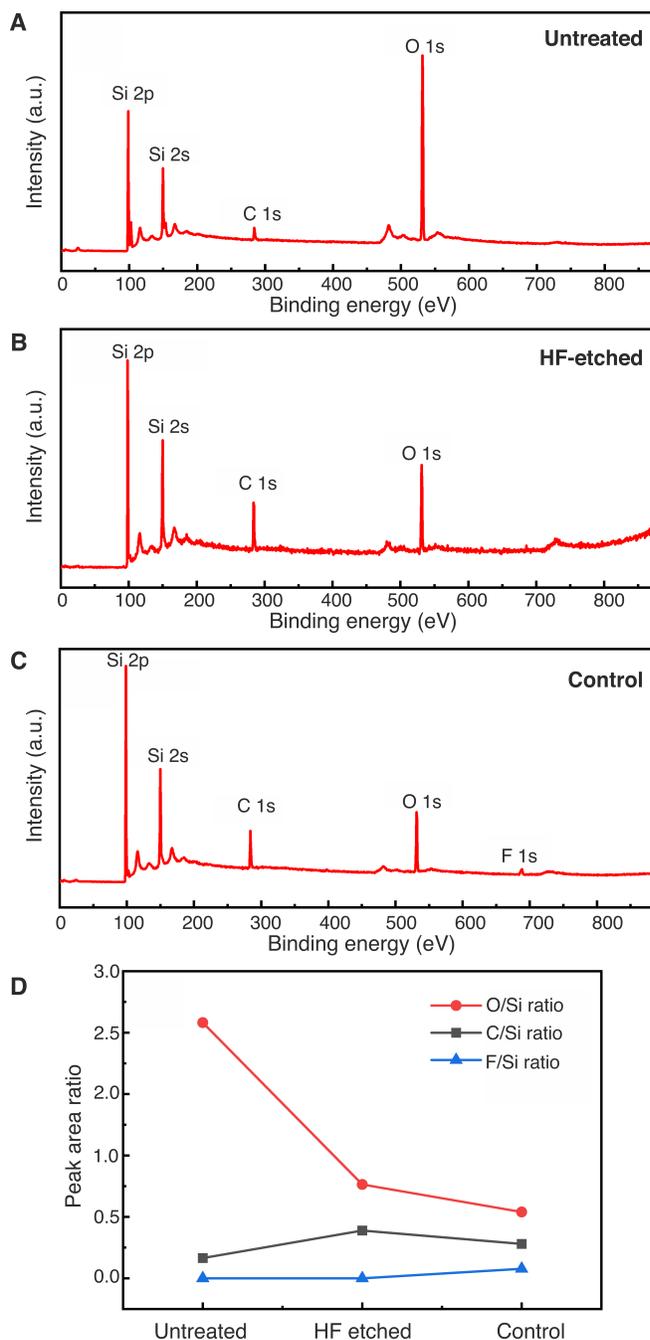

**Fig. 3. XPS survey scans of different Si (100) surfaces.** **(A)** Untreated Si with its native oxide layer. **(B)** As-prepared HF-treated Si. **(C)** Unexposed part of EUV exposed Si (control). **(D)** Peak-area ratio of C, O, and F over Si, for the three different cases, calculated from the corresponding survey scans.

The XPS survey scans of EUV-exposed HF-treated Si surfaces are shown in Fig. 4A-C. In all the samples, the same elements as for the control sample are detected. Figure 4D shows the relative composition of C, F, and O as a function of the exposure dose. Since there is no notable increase in the signal intensity of C with increasing EUV dose, we rule out C deposition during the EUV exposure as the mechanism of resistless patterning. The amount of F is also not significantly affected by the EUV dose, and therefore we conclude that it does not play a role either in the resistless patterning. On the contrary, as the EUV dose is increased, we observe a substantial increase in the oxygen signal. This points towards an oxide formation mechanism. In Fig. 4D we also observe a drop in the O/Si ratio and the C/Si ratio between the control sample (zero dose) and the first dose (6.8 J/cm$^2$). We attribute this to EUV-induced C contamination removal from the Si substrates in the presence of oxygen (*35*).

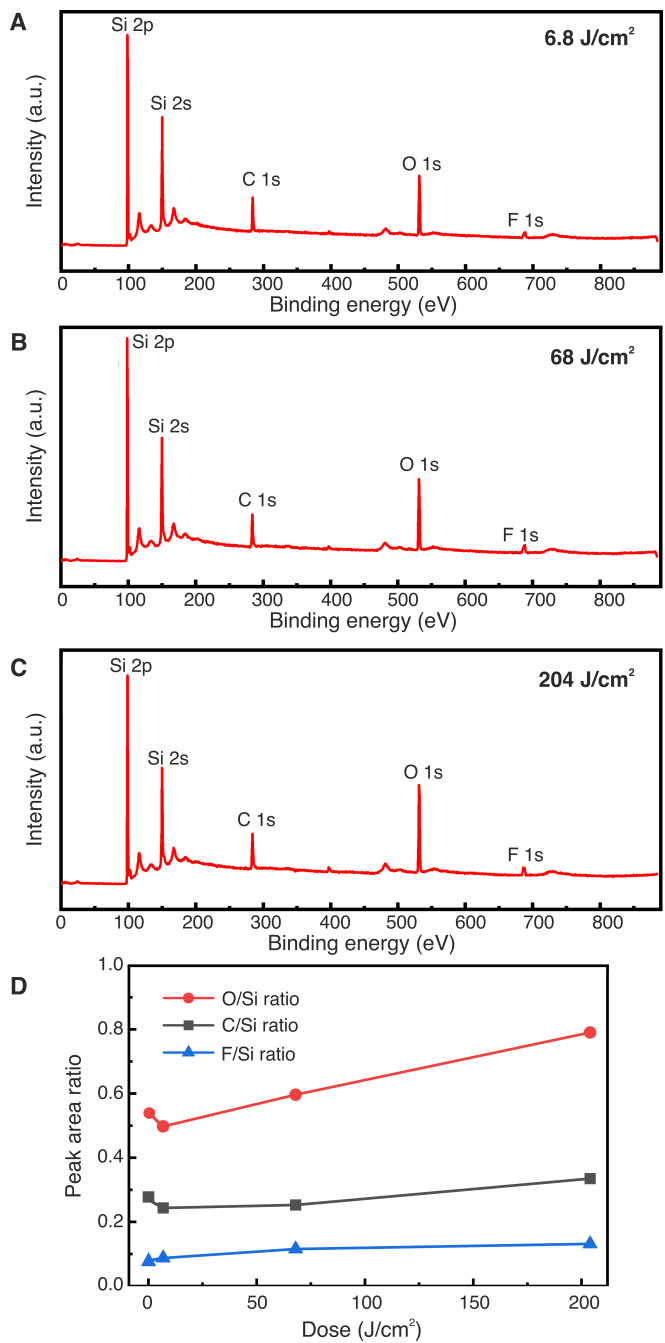

**Fig 4. XPS survey scans of EUV-exposed Si (100) surfaces with different doses. (A)** Low dose of 6.8 J/cm$^2$. **(B)** Medium dose of 68 J/cm$^2$. **(C)** High dose of 204 J/cm$^2$. **(D)** Peak-area ratio of C, O, and F over Si as a function of EUV dose including the control sample at zero dose.

To gain more insight into the EUV-induced surface oxidation, we used a photon energy of 150 eV for the photoelectron excitation during the XPS measurements to reveal the Si 2p peaks associated with Si suboxides, which are difficult to see at the incident photon energy of 990 eV, used for the survey scan. Figures 5A and 5B show the Si 2p core-level spectra acquired from untreated and as-prepared HF-treated Si (100) surfaces, respectively, which correspond to the survey scans in Fig. 3A and 3B (unexposed wafers). The spectra are composed of several components, i.e. the bulk Si 2p, suboxides ($Si^{1+}$, $Si^{2+}$, and $Si^{3+}$), and $SiO_2$ ($Si^{4+}$) and plotted as a function of relative binding energy with respect to bulk Si $2p_{3/2}$ peak in Fig. 5. All the peaks are fitted by keeping the spin-orbit splitting of 0.6 eV and area ratio of 1:2 to assure the physical correctness of the fitting (*36-38*).

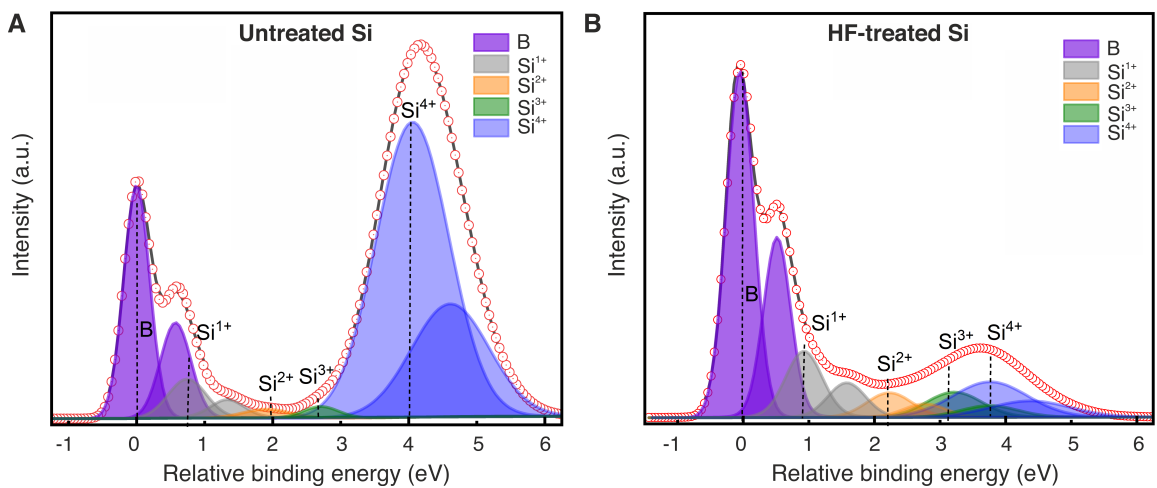

**Fig. 5. Si 2p surface core-level spectra of different unexposed Si (100) surfaces**. **(A)** Untreated Si. **(B)** As-prepared HF-treated Si.

For the untreated Si, the two components of the bulk Si, Si $2p_{1/2}$, and Si $2p_{3/2}$ are designated as B and the corresponding oxides with doublets are observed at relative binding energies 0.75, 1.91, 2.7, and 4.1 eV. We note that a strong peak of $Si^{4+}$ is observed in the Si 2p spectra of untreated Si, which is attributed to the thin layer of native oxide on Si. On the contrary, the $Si^{4+}$ peak is significantly reduced after HF treatment, shown in Fig. 5B. However, the existence of the $Si^{4+}$ peak indicates that the Si surface gets slightly oxidized during its transfer from the HF bath to the vacuum chamber.

A detailed analysis of the Si 2p peak of the samples exposed with different EUV doses is shown in Fig. 6. Fitted Si 2p spectra of the samples exposed with EUV doses 6.8, 68, and 204 $J/cm^2$ are presented in Figs. 6A, B, and C, respectively. As for Fig. 5, the spectra are composed of bulk Si 2p, suboxides ($Si^{1+}$, $Si^{2+}$, and $Si^{3+}$), and $SiO_2$ ($Si^{4+}$). The summed area ratios of all the oxides $SiO_x$ to Si ($SiO_x$/Si) at different EUV doses are plotted in Fig. 6D. It is observed that the ratio of $SiO_x$/Si increases with increasing EUV doses, which reveals that EUV exposure promotes the oxidation of Si. The variation of area ratio of individual suboxides $Si^{1+}$, $Si^{2+}$, $Si^{3+}$, and $Si^{4+}$ to Si at different EUV doses is shown in Fig. 6E. The area ratio of $Si^{1+}$/Si decreases slightly with increasing EUV dose, which implies that $Si^{1+}$ is reduced upon EUV exposure. The area ratio of $Si^{2+}$/Si and $Si^{3+}$/Si increases significantly at 68 $J/cm^2$ and then the ratio is saturated. This reveals that EUV exposure promotes the formation of $Si^{2+}$ and $Si^{3+}$ only up to this dose. On the other hand, the area ratio $Si^{4+}$/Si enhances remarkably beyond 68 $J/cm^2$, which suggests that EUV exposure above this dose results in an increased number of secondary electrons that promote the formation of homogeneous $Si^{4+}$ ($SiO_2$). These results are in accordance with the negative-tone resist behavior of this resistless patterning method and strongly suggest that an increase in EUV dose promotes the formation of stable oxides, more importantly, $SiO_2$ instead of Si suboxides, resulting in a high etch resistance in TMAH-based solutions and thereby leading to an increased etch depth in Si. Indeed, our observation agrees well with the contrast curve of Fig. 2C for E=91.9 eV. At the lowest dose, we do not have enough stable oxide formed and its selectivity to Si in the etching solution is very poor. This is why we have no measured thickness in the contrast curve of Fig. 2C. At the intermediate dose, more $SiO_x$ is formed (increase in the more stable $SiO^{2+}$ and $SiO^{3+}$ and decrease in the less stable $SiO^{1+}$) and therefore some selectivity between the oxide and the Si is observed (about 15 nm of etched depth measured). At the highest dose, $SiO_2$ is clearly formed which exhibits higher selectivity to Si in the TMAH solution and therefore leads to a higher etch depth of about 45 nm. It is remarkable to see that a subtle change of the HF-treated Si surface, upon EUV exposure, can result in the formation of stable oxide, which can serve as an etch mask to pattern Si.

Measuring the thickness of this oxide layer is challenging because it is very thin. To estimate its thickness, we have compared the $Si^{4+}$ peaks of Fig. 5A (native oxide) and Fig. 5A-C (EUV-induced oxide), and in this way we find the oxide to be even thinner than the native oxide. Another indication of the thickness is the time it takes to fully remove this oxide by etching. To fully remove the native oxide by etching in TMAH required approximately 60 s, while to fully remove the EUV-induced oxide required only 5 s (for an exposure dose of 150 J/cm$^2$). Therefore, it is reasonable to infer that the thickness of grown oxide is a nanometer or below.

**High-resolution resistless patterning using EUV achromatic Talbot lithography**

Having demonstrated the capability to oxidise Si with EUV exposure and shed light on the $SiO_x$ formation mechanism induced by the EUV exposure, we set out to determine its prospects for resistless lithography by patterning dense line/space patterns. These were achieved by EUV achromatic Talbot lithography (24). A schematic of the technique is shown in Fig. 7A. Briefly, synchrotron EUV light at an energy of 91.9 eV, which corresponds to a wavelength of 13.5 nm (industry standard), passes through a transmission diffraction grating, fabricated on a thin silicon nitride membrane and placed at a certain distance from the exposed wafer, where a stationary aerial image is formed with 2× demagnification. The stationary intensity is due to the 4% bandwidth of the light, which causes the Talbot self-images of the grating to smear out and eventually merge after a certain distance. The mask used in these experiments consists of a linear grating with a pitch of 300 nm, which corresponds to half-pitch 75 nm lines on the wafer.

Figure 7B shows the top-view scanning electron microscope (SEM) image of half-pitch 75 nm $SiO_2$/Si lines obtained from the EUV ATL exposure with a dose on the mask of roughly 33 J/cm$^2$ after etching in a 10 wt% TMAH + 30 vol% IPA solution at 85 °C for 5 s. The corresponding atomic force microscopy (AFM) measurements in Fig. 7E indicate that the average height of the etched $SiO_2$/Si lines is 31.2 nm. Despite the fact that the EUV achromatic Talbot technique is an efficient process and almost all the transmitted diffraction orders contribute to the aerial image, only 42% of the light is transmitted through the silicon nitride membrane. Therefore, the actual dose on the wafer is significantly lower than the dose on the mask. However, an exact calculation of the dose on the wafer is not necessary and is beyond the scope of this work. In our efforts to further reduce the dose, we found that the dose on the mask could indeed be lowered, and the oxide pattern still developed, by also reducing the etching temperature during the post-exposure treatment. Figures 7C and 7F show the SEM and AFM images respectively of 75 nm $SiO_2$/Si lines etched at 75 °C for 5 seconds after exposure at a dose of about 28 J/cm$^2$. In this case, the average height of the $SiO_2$/Si lines is reduced to 15.8 nm. We could further reduce the dose to 10 J/cm$^2$ by lowering the etching temperature even further, down to 65 °C, as shown in the SEM and AFM images of Figs. 7D and 7G. However, this also further reduced the average etch depth to 11.2 nm.

The obtained SiO$_2$/Si lines are relatively rough in all three cases, which is a result of the wet TMAH etching process and perhaps the H$_2$ bubbles created during its action (*34, 39*). In addition, long exposure times can blur the aerial image due to mechanical drifts between the mask and wafer, which, in turn, will reduce the aerial image contrast and can lead to increased roughness at the nanoscale. This is, however, the smallest half-pitch achieved to date, using top-down photolithography, without the use of a photoresist.

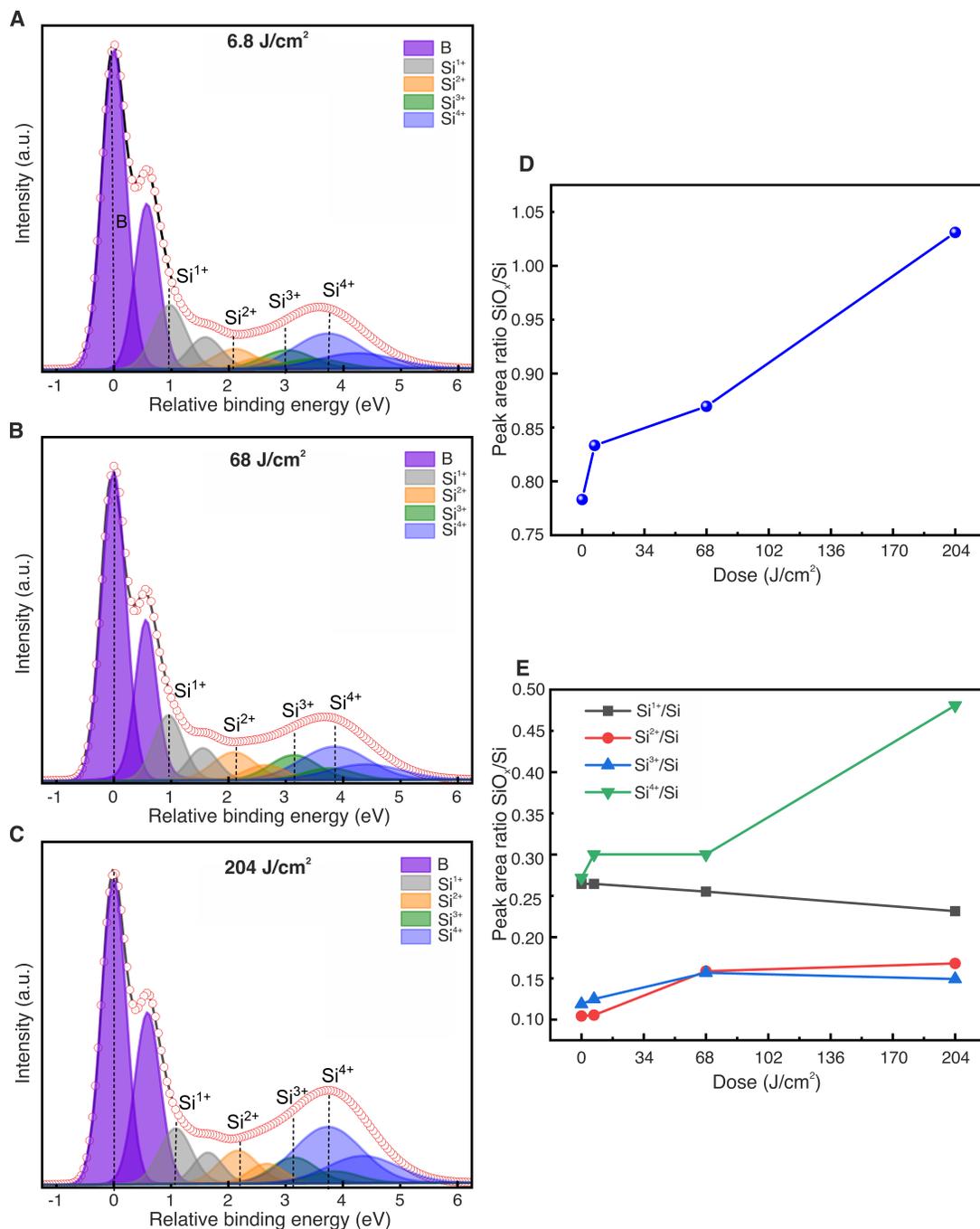

**Fig. 6. Si 2p surface core-level spectra of different Si (100) surfaces exposed with three EUV doses**. (**A**) A low dose of 6.8 J/cm$^2$. (**B**) A medium dose of 68 J/cm$^2$. (**C**) A high dose of 204 J/cm$^2$. (**D**) Peak area ratio of SiO$_x$ to Si. (**E**) Peak area ratio of Si$^{1+}$, Si$^{2+}$, Si$^{3+}$, and Si$^{4+}$ to Si.

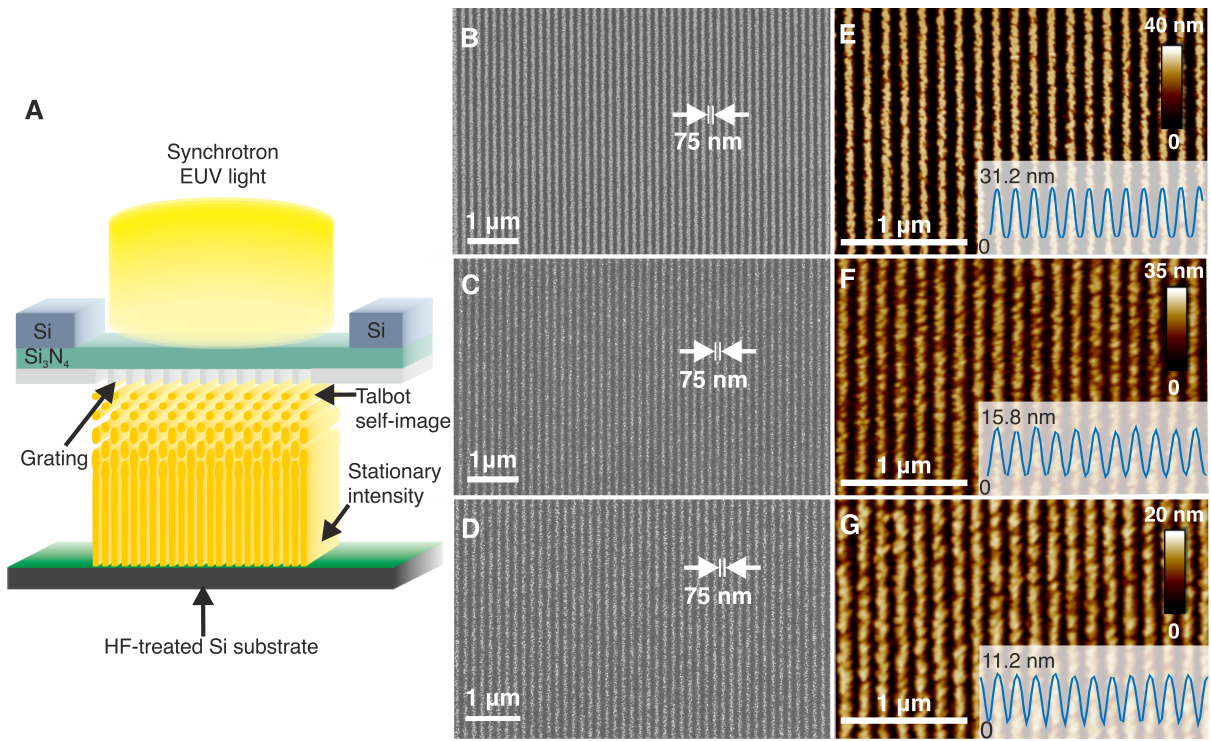

**Fig. 7. High-resolution resistless patterning with EUV achromatic Talbot lithography. (A)** Schematic of the EUV achromatic Talbot lithography technique. **(B)-(D)** Top-view SEM and **(E)-(F)** AFM images of 75 nm $SiO_2$/Si lines patterned on HF-treated Si for three different TMAH solution temperatures: **(B)** and **(E)** 85 °C. **(C)** and **(F)** 75 °C. **(D)** and **(G)** 65 °C. The insets of the AFM images show the average height profile of the $SiO_2$/Si lines.

## Discussion

We have successfully demonstrated nanoscale patterning on HF-treated Si surfaces without the use of a photoresist by EUV exposure and TMAH etching. Furthermore, we have shown that the EUV exposure leads to the formation of a stable oxide layer in the exposed areas. As the EUV dose is increased, more $SiO_x$ is created and especially in its more stable $SiO_2$ form, which results in higher etch resistance in the TMAH solution. We also observed that switching to higher photon energies, above the Si absorption edge, increases the sensitivity as a result of the higher number of photoelectrons due to the increased absorption, Si 2p core-level excitation, and LVV Auger relaxation. Moreover, we have demonstrated, for the first time, dense line/space patterns with half-pitch 75 nm in Si, using EUV achromatic Talbot lithography and TMAH etching.

The obtained nanopattern exhibited relatively high roughness, which can be attributed to multiple effects. The TMAH etching is not ideal for high-resolution patterning, since it is an anisotropic wet etching technique. Development of dry etching methods should improve the roughness (*40, 41*) and resolution (*42*). Moreover, in-situ preparation of the surface will reduce or eliminate its partial oxidation and therefore enhance the contrast of the surface. In particular, combination of this method with surface selective ALD or atomic layer etching (ALE) (*40, 41*) or hydrogen depassivation lithography should substantially enhance the pattern transfer contrast and keep the resolution defined by the exposure. In addition, the relative low sensitivity of the method leads to the extremely long exposure times and therefore the resolution and roughness is limited by the thermal and mechanical stability of the interferometric tool, which will be improved in the future.

With respect to sensitivity, the required doses for the presented patterning mechanism are much higher, compared to the doses required for state-of-the-art EUV resists (i.e. below 100 mJ/cm$^2$). Therefore, the current process is not compatible for high-volume manufacturing of semiconductor devices. The present work should be seen as a proof of principle, demonstrating resistless patterning with EUV light. Firstly, we have shown that EUV photons can induce solubility switch in inorganic materials, whereas past and current research and development of EUV resists has focused on organic materials (chemically amplified resists) and recently on metal-organic resists. Secondly, surface modification or resists of monolayer thickness can be sufficient for pattern transfer. Although the current approach is very insensitive, we expect it to spark further studies to develop inorganic materials and surface monolayers with increased EUV absorption and faster solubility switch. Along this direction, we plan to investigate other Si surfaces such as native $SiO_2$ and piranha-etched. Our preliminary experiments have shown that these surfaces require lower doses, leading to better results; these experiments are ongoing and will be the subject of a future publication. Performing patterning experiments under controlled atmosphere of water and oxygen will also provide a means to improve the process and lower the required photon doses.

While there is still a lot of work to be done before this technique is suitable for industrial integration, we are enthusiastic about its potential for high-resolution patterning and there are many promising avenues for future research. Further development and new approaches of resistless EUV lithography could open a new way for device fabrication, enabling large-area patterning of high-resolution features compatible simultaneously with the high-volume manufacturing of ICs and atomic-scale fabrication methods but without the limitations imposed by photoresists. Additionally, combining this technique with selective area doping or ALD could enable the wafer-scale patterning of interconnects between the active components of quantum devices (*9-12*), facilitating the scale-up of these devices to large numbers.

## Materials and Methods

Hydrogen-terminated Si (100) surfaces are prepared by immersing a clean Si (100) 4" wafer into a buffered HF solution (Buffered Oxide Etch 7:1, Technic, France) for 5 min, followed by a quick rinse of deionized (DI) water (< 10 s). We then pack the Si wafer in a nitrogen atmosphere before loading it into the EUV chamber. We carry out the EUV exposures at the XIL-II beamline of the Swiss Light Source (SLS) with a chamber pressure of ~$5 \times 10^{-7}$ mbar. The beamline and the EUV interference lithography endstation are described in detail elsewhere (*26*). The dose tests are done by open-frame exposures with an aperture of 0.5×0.5 mm$^2$ at a wide range of doses at different photon energies of 107.8 eV (λ=11.5 nm), 102 eV (λ=12.15 nm), and 91.9 eV (λ=13.5 nm). After the EUV exposures, the Si wafers are taken out of the EUV vacuum chamber and packed in $N_2$ ambient before any post-exposure treatment. The exposed areas are developed by using three different solutions for dose optimization: 25 wt% TMAH (VLSI grade, Technic, France), 10 wt% TMAH, and 10 wt% TMAH + 30 vol% IPA (ULSI grade, Technic, France) at 85 °C for 5 s followed by a quick DI water rinse and $N_2$ blow-drying. We then measure the obtained etched Si depth with a profilometer (Veeco, Dektak 150), plot it as a function of the exposure dose, and fit it with a dose-response curve. The line/space patterns are exposed using EUV achromatic Talbot lithography with a 300 nm-pitch mask grating using a photon energy of 91.9 eV. The resulting aerial image recorded on the substrate has a pitch of 150 nm, due to the two times demagnification of the EUV achromatic Talbot lithography technique. After the EUV exposure, the Si wafer is etched in 10% TMAH + 30 vol% IPA at 65, 75, and 85 °C. The top-view images of the etched $SiO_2$/Si lines are obtained by SEM (Zeiss Supra 55 VP) and AFM (Bruker Dimension 3100) imaging, respectively.

We use synchrotron radiation XPS to investigate the changes in the chemical bonding configurations on the Si surface before and after the EUV exposures. The measurements are carried out at the PEARL beamline of the SLS (*43*). We carry out the survey scans with an incident photon energy of 990 eV, an incidence angle of 30º, and a beam spot size of 190×70 µm$^2$. To enhance the surface signal sensitivity, we use a photon energy of 150 eV and an incidence angle of 30º to measure the Si 2p spectra with a pass energy of 10 eV. For all the experiments, the take-off angle is perpendicular to the sample surface.

**Acknowledgments**

The authors thank Dr. Iacopo Mochi, Michaela Vockenhuber, and Markus Kropf for their technical support at the XIL-II beamline in SLS. They are also grateful to Dr. Nicolas Bachellier for the fruitful discussions and technical assistance at the PEARL beamline.

**Funding**

The project was financially supported by the Engineering and Physical Sciences Research Council (EPSRC) project EP/M009564/1. P.C.C. was partly supported by the EPSRC Centre for Doctoral Training in Advanced Characterization of Materials (grant number EP/L015277/1), by the Paul Scherrer Institute and also the Microsoft Corporation.

**Author contributions:**
Conceptualization: GA, YE, LTT, DK
Validation: GA, YE, DK, LTT, PK, PCC, TJZS, NJC, SRS, MM
Methodology: GA, YE, LTT, DK, PCC
Investigation: LTT, MM
Visualization: LTT, PK, DK
Supervision: YE, DK
Writing—original draft: LTT, PK, DK, YE
Writing—review & editing: DK, PK, LTT, GA, YE, PCC, TJZS, NJC, SRS, MM

**Competing interests:** The authors declare they have no competing interest.

**Data and materials availability:** All data needed to evaluate the conclusions in the paper are present in the paper.